\documentclass[11pt]{article}
\usepackage{verbatim}
\usepackage{amsmath,amssymb}
\makeatletter \@addtoreset{equation}{section} \makeatother

\newcommand{\noi}{\vspace{12pt}\noindent}
\newcommand{\beq}{\begin{equation}}
\newcommand{\eeq}{\end{equation}}
\newcommand{\bea}{\begin{eqnarray}}
\newcommand{\eea}{\end{eqnarray}}

\newcommand{\e}[1]{{(\ref{#1})}}
\newcommand{\eq}[1]{{eq.\ (\ref{#1})}}
\newcommand{\es}[2]{{(\ref{#1}) and (\ref{#2})}}

\newcommand{\Ref}[1]{{Ref.~\cite{#1}}}

\newcommand{\equi}[1]{\stackrel{{#1}}{=}}

\newcommand{\Z}{\mathbb{Z}}
\newcommand{\sgn}{\mathrm{sgn}}

\newcommand{\eg}{{${ e.g., \ }$}}

\newcommand{\cf}{{cf.\ }}

\newcommand{\wrt}{{with respect to }}

\newcommand{\wtho}{{with the help of }}

\newcommand{\lhs}{{left-hand side }}

\renewcommand{\~}{ \ }
\renewcommand{\=}{ \ = \ }
\newcommand{\+}{ \ + \ }
\renewcommand{\-}{ \ - \ }

\newcommand{\q}{{}^{}}

\newcommand{\for}{\mathrm{for}}

\newcommand{\Hf}{{\frac{1}{2}}}

\newcommand{\proofbox}{\begin{flushright}{\hfill \ensuremath{\Box}}
\end{flushright}}

\newtheorem{theorem}{Theorem}[section]

\begin{document}
\thispagestyle{empty}
\title{\Large{\bf A Note on Angular Momentum Commutators \\
in Light-Cone Formulation of \\ Open Bosonic String Theory}}
\author{{\sc Klaus~Bering}$^{a}$ \\~\\
$^{a}$Institute for Theoretical Physics \& Astrophysics\\
Masaryk University\\Kotl\'a\v{r}sk\'a 2\\CZ--611 37 Brno\\Czech Republic}

\maketitle
\vfill

\begin{abstract} 
We recalculate in a systematic and pedagogical way one of the most important
results of Bosonic open string theory in the light-cone formulation, namely the
$[J^{-i},J^{-j}]$ commutators, which together with Lorentz covariance, famously
yield the critical dimension $D=26$ and the normal order constant $a=1$. 
We use traditional transverse oscillator mode expansions (avoiding the
elegant but more advanced language of operator product expansions). We
streamline the proof by introducing a novel bookkeeping/regularization
parameter $\kappa$ to avoid splitting into creation and annihilation parts,
and to avoid sandwiching between bras and kets.
\end{abstract}

\vfill

\begin{quote}
PACS number(s): 04.60.Ds; 11.25.-w; 11.25.Hf; 11.30.Cp.  
\hrule width 5.cm \vskip 2.mm \noindent 
$^{a}${\small E--mail:~{\tt bering@physics.muni.cz}} \\
\end{quote}

\newpage

\section{Introduction}

\noi
One of the most fundamental facts of Bosonic string theory, that a student of
string theory would want to rederive for himself, is the critical dimension
$D=26$. There are traditionally four\footnote{Plus various heuristic
arguments typically involving zeta function regularization \cite{gsw87,bz04}.}
ways to detect the critical dimension $D=26$ in Bosonic string theory. 
\begin{enumerate}

\item
Preservation of Lorentz covariance in the light-cone formulation
\cite{grt72,ggrt73}.

\item
No negative norm states/ghosts in the covariant formulation \cite{gt72}.  

\item
The vanishing of the conformal/Weyl anomaly in Polyakov's path integral 
formulation \cite{polyakov81}.

\item
Nilpotency of the BRST generator in the covariant formulation \cite{ko83}. 
\end{enumerate}

\noi 
Here we will only consider the first method in the open string case in a flat 
Minkowski target space. Students of Zwiebach's book \cite{bz04}, which uses the
light-cone formulation, will notice that the book in Section 12.5 stops short
of proving\footnote{\Ref{gsw87} gives a proof by sandwiching $[J^{-i},J^{-j}]$
between a bra and a ket, which depend on the choice of vacuum.}
the critical dimension $D=26$. The goal of the current paper is
to fill that gap in a pedagogical and efficient manner. The main calculation is
an evaluation of a commutator $[E^{i},E^{j}]$ between two expressions $E^{i}$
and $E^{j}$, which are cubic in the transverse $\alpha$ oscillator modes, see
Section~\ref{secee}. The original papers of Goddard, Goldstone, Rebbi and
Thorn \cite{grt72,ggrt73}, who splits in creation and annihilation parts,
are sparse on details, although recently an explicit calculation has appeared
in \Ref{arutyunov09} for the closed string case. Here we will use a more
efficient method by introducing a $\kappa$ regularization parameter
\cite{bb06}, see Section~\ref{seckappa}, so that we can rigorously calculate
with symmetrized expressions via Wick's Theorem \cite{wick50}, see
Section~\ref{secwick}. We believe the techniques displayed here are interesting
in their own right, applicable far beyond the shown calculations.

\section{Basic Settings}

\noi
To keep this paper short, it is necessary to assume that the reader is
familiar with the light-cone formulation of Bosonic string theory. Here we will
only briefly repeat all relevant definitions and formulas to set notations and
conventions. {}For explanations and justifications, we defer to, \eg
Zwiebach's book \cite{bz04}.

\noi
The light-cone metric in flat Minkowski target space is
\beq
\eta\q_{\mu\nu}\~\equiv\~\left[\begin{array}{cc|ccc} 0&-1&0&0&\cdots \cr
-1&0&0&0&\cdots  \cr\hline 0&0&1&0&\cdots  \cr  0&0&0&1& \cr
\vdots&\vdots&\vdots&&\ddots \end{array} \right] \~, \qquad
\mu,\nu\~\in\~\{+,-,i\}\~, 
\qquad i\~\in\~\{1,\ldots, D\!-\!2\}\~. \label{etametric}
\eeq
Here Greek indices $\mu,\nu,\ldots,$ runs over all target space dimensions, 
while Latin indices $i,j,\ldots,$ only runs over transversal directions.
We normalize the center-of-mass position $x^{\mu}_{0}$ and the total momentum 
$p^{\mu}$ of the open string as follows\footnote{
Note that Goddard, Goldstone, Rebbi and Thorn \cite{ggrt73} 
call the center-of-mass position $x^{\mu}_{0}$ for $q^{\mu}_{0}$.}
\beq 
q^{\mu}_{0} \~\equiv\~ \frac{x^{\mu}_{0}}{c\sqrt{2\hbar\alpha^{\prime}}}
\~\equiv\~ \sqrt{\frac{\hbar}{2}} \frac{x^{\mu}_{0}}{\ell_{s}}
\~, \qquad
\alpha^{\mu}_{0} \~\equiv\~ c\sqrt{2\hbar\alpha^{\prime}}p^{\mu}
\~\equiv\~ \sqrt{\frac{2}{\hbar}}\ell_{s}p^{\mu}\~,  \qquad
\alpha^{\prime}\~\equiv\~\frac{1}{2\pi\hbar c T_{0}}\~,  \qquad 
\ell\q_{s}\~\equiv\~\hbar c\sqrt{\alpha^{\prime}}\~,\label{qxpalf}
\eeq
where the string tension $T_{0}$ has dimension of force, and 
$1/\sqrt{\alpha^{\prime}}$ has dimension of energy. The fundamental dynamical
operators in the light-cone formalism (in light-cone gauge) are
\beq
q^{-}_{0}\~, \qquad  \alpha^{+}_{0}\~, \qquad q^{i}_{0}\~, \qquad 
\alpha^{i}_{n}\~, \label{fundynops}
\eeq
where $n\in\Z$ and $i=1,\ldots, D-2$. All the operators \e{fundynops} have the
same dimension as $\sqrt{\hbar}$. The notations for the commutator and the
anti-commutator of two operators $A$ and $B$ are
\beq
[A,B]\~\equiv\~AB-BA \~, \qquad\qquad \{A,B\}\~\equiv\~AB+BA\~,
\eeq
respectively. More generally, define the $n$-symmetrizer
\beq
\{A\q_{1},\ldots, A\q_{n}\}
\~\equiv\~ \sum_{\pi\in S\q_{n}} A\q_{\pi(1)}\ldots A\q_{\pi(n)}
\label{nsymmetrizer}
\eeq 
of $n$ operators $A_{1}, \ldots, A_{n}$, as sum of all possible permutations 
$\pi\in S\q_{n}$. Normal ordering (usually denoted with a double colon) moves
all the annihilation operators $\alpha^{i}_{m>0}$ to the right of all the
creation operators $\alpha^{i}_{m<0}$. Equivalently in formula,
\beq
:\alpha^{i}_{m}\alpha^{j}_{n} :
\~\=\~\theta(n\!-\!m)\alpha^{i}_{m}\alpha^{j}_{n} 
\+ \theta(m\!-\!n)\alpha^{j}_{n}\alpha^{i}_{m}\~, \label{normalorder} 
\eeq
where $\theta$ denotes the Heaviside step function with $\theta(0)=\Hf$. Note
that operators commute inside the normal order symbol, for instance
$:AB:\~\=\~:BA:$ for two operators $A$ and $B$.

\section{Fundamental Commutator Relations and the $\kappa$ Parameter}
\label{seckappa}

\noi
The non-zero commutator relations for the fundamental operators \e{fundynops}
are\footnote{
An ``$\mathrm{i}$'' that is not an upper or lower index does always 
denote the imaginary unit.}
\beq
[q^{-}_{0},\alpha^{+}_{0}]\= \mathrm{i}\hbar\eta^{-+}
\~\equi{\e{etametric}}\~ -\mathrm{i}\hbar\~,
\qquad
[q^{i}_{0},\alpha^{j}_{0}] \= \mathrm{i}\hbar\eta^{ij}\~,  \qquad 
[\alpha^{i}_{m},\alpha^{j}_{n}] \= \hbar m\kappa^{|m|}\delta^{0}_{m+n}\eta^{ij}\~.
\label{ccr01}
\eeq
The parameter $\kappa$ in the commutator relation \e{ccr01} is a regularization
parameter with $|\kappa|<1$. In the end of the calculations, one should take 
the limit $\kappa\to 1$. The limit $\kappa\to0$ corresponds to the classical
limit $\hbar\to0$. Now why do we introduce the regularization parameter 
$\kappa$? To answer this question, imagine in the standard $\kappa=1$ case, 
that we want to calculate the commutator $C=[A,B]$ of two normal-ordered
operators $A$ and $B$ (which are polynomials in the $\alpha$ oscillator modes)
by carefully performing a minimal number of $\alpha\alpha$ commutations to
bring $C=[A,B]$ on normal-ordered form. Imagine further that the result $C$
happens to be finite. Then convergence can only improve if we repeat the
calculation $C(\kappa)=[A,B]$ with $|\kappa|<1$. Moreover, the result will
depend continuously $C(\kappa)\to C(1)$ as $\kappa\to 1$. This suggests a
strategy. We first introduce the regularization parameter $\kappa$ in the
commutator relation \e{ccr01} with $|\kappa|<1$. As we shall soon see, the
commutator $C(\kappa)=[A,B]$ will remain well-defined under a wider and more
powerful class of mathematical manipulations as long as $|\kappa|<1$. Thus we
can calculate the commutator $C(\kappa)$ more efficiently, and in the end, we 
take the limit $\kappa\to 1$.

\section{Wick's Theorem}
\label{secwick}

\noi
We list here some of the first few consequences of Wick's Theorem
\cite{wick50}, which will be our main computational tool.

\begin{theorem}[Wick's Theorem for symmetrization and normal order]
i) The anti-commutator 
\beq
\Hf\{\alpha^{i}_{m},\alpha^{j}_{n}\}
\~\=\~:\alpha^{i}_{m}\alpha^{j}_{n}: \+ c^{ij}_{mn}\~, \qquad\qquad
c^{ij}_{mn}\~\equiv\~\frac{\hbar}{2}|m|\kappa^{|m|}\delta^{0}_{m+n}\eta^{ij}\~, 
\label{wsn1}
\eeq
is a sum of a normal ordered term and a single contraction term.
ii) The $4$-symmetrizer
\bea
\frac{1}{24}\{\alpha^{i_{1}}_{n_{1}},\alpha^{i_{2}}_{n_{2}},\alpha^{i_{3}}_{n_{3}},
\alpha^{i_{4}}_{n_{4}}\}
&=&:\alpha^{i_{1}}_{n_{1}}\alpha^{i_{2}}_{n_{2}}
\alpha^{i_{3}}_{n_{3}}\alpha^{i_{4}}_{n_{4}}: 
\+\frac{6}{24}\sum_{\pi\in S_{4}}c^{i_{\pi(1)}i_{\pi(2)}}_{n_{\pi(1)}n_{\pi(2)}}
:\alpha^{i_{\pi(3)}}_{n_{\pi(3)}}\alpha^{i_{\pi(4)}}_{n_{\pi(4)}}: 
\cr
&&\+\frac{3}{24}\sum_{\pi\in S_{4}}
c^{i_{\pi(1)}i_{\pi(2)}}_{n_{\pi(1)}n_{\pi(2)}}c^{i_{\pi(3)}i_{\pi(4)}}_{n_{\pi(3)}n_{\pi(4)}}
\label{wsn2}
\eea
is a sum of a normal ordered term, $6$ different single contraction terms 
and $3$ different double contraction terms.
\label{wickthm1}
\end{theorem}

\begin{theorem}[Wick's Theorem for commutators and symmetrization]
i) If $[A\q_{a},B\q_{b}]$ are $c$-numbers, $a,b=1,2 \mod 2$, then the
commutator of anti-commutators
\beq
\Hf\left[\Hf\{A\q_{1},A\q_{2}\},\Hf\{B\q_{1},B\q_{2}\} \right]
\=\sum_{a,b=1}^{2} \Hf[A\q_{a},B\q_{b}]\Hf\{A\q_{a+1},B\q_{b+1}\}
\label{wcs2}
\eeq
is a sum of single commutator terms. 
ii) If $[A\q_{a},B\q_{b}]$ are $c$-numbers, $a,b=1,2,3 \mod 3$, then the
commutator of $3$-symmetrizers
\bea
\Hf\left[\frac{1}{6}\{A\q_{1},A\q_{2},A\q_{3}\},
\frac{1}{6}\{B\q_{1},B\q_{2},B\q_{3}\}\right]
&=&\sum_{a,b=1}^{3}\Hf[A\q_{a},B\q_{b}]
\frac{1}{24}\{A\q_{a+1},A\q_{a+2},B\q_{b+1},B\q_{b+2}\} \cr
&&+\sum_{\pi\in S\q_{3}}\Hf[A\q_{1},B\q_{\pi(1)}]
\Hf[A\q_{2},B\q_{\pi(2)}]\Hf[A\q_{3},B\q_{\pi(3)}]\~.\label{wcs3}
\eea
is a sum of single and triple commutator terms.
\label{wickthm2}
\end{theorem}

\noi
Wick's Theorem~\ref{wickthm2} follows by expanding out to appropriate order 
the ``little Baker-Campbell-Hausdorff formula'' $e^{A}e^{B}=e^{A+B+\Hf[A,B]}$
(which holds if $[A,B]$ is a $c$-number)
with $A=\sum_{a}x^{a}A\q_{a}$ and $B=\sum_{b}y^{b}B\q_{b}$, where $x^{a}$ and
$y^{b}$ are parameters, and then afterwards antisymmetrize \wrt
$A\leftrightarrow B$ on both sides.

\section{Transverse Virasoro Generators $L^{\perp}_{n}$ and Algebra}

\noi
The $\alpha^{-}_{n}$ modes and the transverse Virasoro generators 
$L^{\perp}_{n}$ are defined as\footnote{Zwiebach \cite{bz04} defines the normal
ordering constant $a$ with the opposite sign.}
\bea
\alpha^{-}_{n}
&\equiv&\frac{1}{\alpha^{+}_{n}}(L^{\perp}_{n}-\hbar a\delta^{0}_{n})\~, 
\qquad n\in\Z\~,\label{alphaminus} \\ 
L^{\perp}_{n}&\equiv&\Hf\eta\q_{ij}\sum_{k\in\Z} :\alpha^{i}_{n-k}\alpha^{j}_{k}:
\~\stackrel{k=\ell+\frac{n}{2}}{=}\~\Hf\eta\q_{ij}\sum_{\ell\in\Z+\frac{n}{2}}
:\alpha^{i}_{\frac{n}{2}-\ell}\alpha^{j}_{\frac{n}{2}+\ell}: \label{lnperpdef01} \\
&\equi{\e{wsn1}}&
\frac{1}{4}\eta\q_{ij}\sum_{k\in\Z}\{\alpha^{i}_{n-k},\alpha^{j}_{k}\}
\- \hbar\frac{D-2}{4}\delta^{0}_{n}\sum_{k\in\Z}|k|\kappa^{|k|}\~.
\label{lnperpdef02}  
\eea
It may at first seem a bit cumbersome to sum over half-integers $\ell$ in
\eq{lnperpdef01}, but it makes the symmetry $\ell\leftrightarrow -\ell$ 
manifest, which is sometimes convenient. Notice that the last $c$-number sum 
\beq
\sum_{k\in\Z}|k|\kappa^{|k|}\=\kappa\frac{d}{d\kappa}\sum_{k\neq 0}\kappa^{|k|}
\~\equi{|\kappa|<1}\~\kappa\frac{d}{d\kappa}\frac{2\kappa}{1-\kappa}
\=\frac{2\kappa}{(1-\kappa)^{2}} \label{simplesum}
\eeq
in \eq{lnperpdef02} is absolutely and unconditionally convergent for 
$|\kappa|<1$ but divergent for $|\kappa|>1$ and $\kappa=1$. (Zeta function 
regularization would suggest that one should assign the value 
$2\sum_{k>0}k\sim 2\zeta(-1)=-\frac{1}{6}$ to the sum \e{simplesum} at 
$\kappa=1$.) We precisely introduced the regularization parameter $\kappa$ to
be able to rearrange expressions without encountering infinities.
The non-zero commutator relations between the transverse Virasoro 
generators $L^{\perp}_{n}$ and the fundamental variables read
\beq
[\alpha^{i}_{m},L^{\perp}_{n}]\=\hbar m\kappa^{|m|}\alpha^{i}_{m+n}\~,
\qquad\qquad  [q^{i}_{0},L^{\perp}_{n}]\= \mathrm{i}\hbar\alpha^{i}_{n}\~.
\label{lnalfm}
\eeq
As a warm-up exercise, let us derive the transverse Virasoro algebra with
central charge $c=D-2$,
\beq
\fbox{$
[L^{\perp}_{m},L^{\perp}_{n}]\= \hbar(m-n)L^{\perp}_{m+n}
+\hbar^{2}\frac{D-2}{12}m(m^{2}-1)\kappa^{|m|}\delta^{0}_{m+n}
+{\cal O}(\kappa\!-\!1)\~.$}\label{viralg}
\eeq
{\sc Proof of \eq{viralg}.} The commutator on the \lhs of \eq{viralg} is a sum
of two terms
\bea
C_{mn}&\equiv&[L^{\perp}_{m},L^{\perp}_{n}]\~\equi{\e{lnperpdef02}}\~
\sum_{k,\ell\in\Z}\left[\frac{1}{4}\{\alpha^{i}_{m-k}, \alpha^{i}_{k}\},
\~\frac{1}{4}\{\alpha^{j}_{n-\ell}, \alpha^{j}_{\ell}\} \right]
\~\equi{\e{wcs2}}\~\sum_{k,\ell\in\Z}[\alpha^{i}_{k},\alpha^{j}_{\ell}]
\Hf\{\alpha^{i}_{m-k},\alpha^{j}_{n-\ell}\} \cr
&\equi{\e{ccr01}}&
\hbar\sum_{k\in\Z} k\kappa^{|k|}\Hf\{\alpha^{i}_{m-k},\alpha^{i}_{n+k}\} 
\~\equi{\e{wsn1}}\~
\hbar\sum_{k\in\Z}k\kappa^{|k|}\left( :\alpha^{i}_{m-k}\alpha^{i}_{n+k}:
\+ c^{ii}_{m-k,n+k} \right)\cr
&=&C^{(2)}_{mn}\+ C^{(0)}_{mn}\~.
\eea
The first term $C^{(2)}_{mn}$ is quadratic (hence the superscript ``$2$'') in 
the transverse $\alpha$ oscillator modes
\bea
C^{(2)}_{mn}&\equiv&\hbar\sum_{k\in\Z}k\kappa^{|k|} 
:\alpha^{i}_{m-k}\alpha^{i}_{n+k}:
\~\stackrel{k=\ell+\frac{m-n}{2}}{=}\~
\hbar\sum_{\ell\in\Z+\frac{m+n}{2}}(\frac{m-n}{2}+\ell)
\kappa^{|\frac{m-n}{2}+\ell|}
:\alpha^{i}_{\frac{m+n}{2}-\ell}\alpha^{i}_{\frac{m+n}{2}+\ell}: \cr
&\stackrel{\ell\leftrightarrow -\ell}{=}&
\frac{\hbar}{2}\sum_{\ell\in\Z+\frac{m+n}{2}}
\left[(\frac{m-n}{2}+\ell)\kappa^{|\frac{m-n}{2}+\ell|}
+(\frac{m-n}{2}-\ell)\kappa^{|\frac{m-n}{2}-\ell|}\right]
:\alpha^{i}_{\frac{m+n}{2}-\ell}\alpha^{i}_{\frac{m+n}{2}+\ell}: \cr 
&\longrightarrow&\hbar(m-n)L^{\perp}_{m+n} \~\~\for\~\~\kappa\to 1\~.
\eea
The second term $C^{(0)}_{mn}$ is the $c$-number anomaly term
\beq
C^{(0)}_{mn}\~\equiv\~\hbar\sum_{k\in\Z}k\kappa^{|k|}c^{ii}_{m-k,n+k}
\~\equi{\e{wsn1}}\~
\frac{\hbar^{2}}{2}\sum_{k\in\Z}k\kappa^{|k|}|m\!-\!k|\kappa^{|m-k|}
\delta^{0}_{m+n}\eta^{ii}
\~\equi{\e{am}}\~\hbar^{2}\frac{D-2}{2}A_{m}\delta^{0}_{m+n}\~,
\eeq
with anomaly
\beq
A\q_{m}\~\equiv\~\sum_{k\in\Z}k|m\!-\!k|\kappa^{|k|+|m-k|}
\~\stackrel{\ell=m-k}{=}\~
\sum_{\scriptsize\begin{array}{c}k,\ell\in\Z \cr k+\ell=m \end{array}}
k|\ell|\kappa^{|k|+|\ell|} 
\=\frac{m(m^{2}-1)}{6}\kappa^{|m|}\~. \label{am}
\eeq
Standard reasoning shows that the $\kappa$ power series \e{am} is absolutely
and unconditionally convergent for $|\kappa|<1$. However, one can say more.
The following argument reveals that the $\kappa$ power series \e{am} only has
one non-zero coefficient, and therefore is just a monomial in $\kappa$, which
makes sense for any $\kappa\in\mathbb{C}$. In the restricted double summation
\e{am}, note that the $(k,\ell)$'th term is antisymmetric under a
$(k \leftrightarrow \ell)$ exchange if the summation variables $k$ and $\ell$
have opposite signs. Therefore one only has to consider $k$'s and $\ell$'s with
weakly the same sign. (The word {\em weakly} refers to that $k$ or $\ell$ could
be $0$.) Since at the same time the sum $k+\ell = m$ of $k$ and $\ell$ is held
fixed, the restricted $(k,\ell)$ double sum contains only finitely many terms,
all with the same power $|m|$ of $\kappa$, and which may be readily summed.
Since $A\q_{-m}=-A\q_{m}$ is odd, it is enough to consider $m\geq 1$. Then 
\beq
A\q_{m}\=\kappa^{m}\sum_{k=1}^{m} k(m-k)
\=\frac{m(m^{2}-1)}{6}\kappa^{m}\~, 
\qquad m\geq 1\~,\label{am2}
\eeq
which, \eg follows from the fact that $\sum_{k=1}^{m}k=\Hf m(m+1)$ and 
$\sum_{k=1}^{m}k^{2}=\frac{1}{3} m(m+\Hf)(m+1)$ for $m\geq 1$.
\proofbox

\noi
By similar arguments, one may derive that the following $\kappa$ power series
\e{bm} is also just a monomial in $\kappa$,
\beq
B\q_{m}\~\equiv\~\sum_{k\in\Z}\sgn(k)\kappa^{|k|+|m-k|}
\~\stackrel{\ell=m-k}{=}\~
\sum_{\scriptsize\begin{array}{c}k,\ell\in\Z \cr k+\ell=m \end{array}}
\sgn(k)\kappa^{|k|+|\ell|} 
\=m\kappa^{|m|}\~, \label{bm}
\eeq
which we will need later in \eq{eesinglenotr}.

\section{Angular Momentum $J^{\mu\nu}$}

\noi
The angular momentum $J^{\mu\nu}$ consists of a center-of-mass part 
$\ell^{\mu\nu}$ and an oscillator part $E^{\mu\nu}$,
\bea
J^{\mu\nu}&\equiv&\ell^{\mu\nu}+E^{\mu\nu}\=-\~(\mu\leftrightarrow\nu)\~,
\label{jaymunu} \\
\ell^{\mu\nu}&\equiv&\Hf \{x^{\mu}_{0},p^{\nu}_{0}\}
\- (\mu\leftrightarrow\nu)
\~\equi{\e{qxpalf}}\~\Hf\{q^{\mu}_{0},\alpha^{\nu}_{0}\}
\- (\mu\leftrightarrow\nu)\~,\label{ellmunu} \\
E^{\mu\nu} &\equiv&-\sum_{n\neq 0}\frac{\mathrm{i}}{n}
:\alpha^{\mu}_{-n} \alpha^{\nu}_{n}:
\~\equi{\e{normalorder}}\~
-\sum_{n>0}\frac{\mathrm{i}}{n}\alpha^{\mu}_{-n}\alpha^{\nu}_{n}
\- (\mu\leftrightarrow\nu)
\~\equi{\e{ccr01}}\~
\sum_{n\neq 0}\frac{\mathrm{i}}{2n}\{\alpha^{\mu}_{-n},\alpha^{\nu}_{n}\}\~,
\label{eemunu}
\eea
where $\mu,\nu\in\{-,i\}$. (Recall that $x^{+}_{0}$ and $J^{\mu+}$ are somewhat
amputated in the light-cone formalism \cite{bz04}.) The angular momentum
$J^{-i}$ consists of three terms\footnote{Conventions differ slightly between
various references,
$\mathrm{i}E^{i}\equiv -\mathrm{i}E^{i}_{\mathrm{GSW}}\equiv 
E^{i}_{\mathrm{GRT}}\equiv E^{i}_{\mathrm{SCH}}\equiv 
-\sum_{n\neq 0}\frac{1}{n}:\alpha^{i}_{-n}L^{\perp}_{n}:$,
and 
$\mathrm{i}E^{ij}\equiv \mathrm{i}E^{ij}_{\mathrm{GSW}}
\equiv E^{ij}_{\mathrm{GRT}}\equiv E^{ij}_{\mathrm{SCH}}\equiv
\sum_{n\neq 0}\frac{1}{n}:\alpha^{i}_{-n} \alpha^{j}_{n}:$,
where GRT$\equiv$\Ref{grt72}, SCH$\equiv$\Ref{schwarz82} and 
GSW$\equiv$\Ref{gsw87}.}
\bea
J^{-i} &\equiv& \ell^{-i}_{I}+\ell^{-i}_{II}+E^{-i}\~, \qquad\qquad  
\ell^{-i}_{I} \~\equiv\~ q^{-}_{0}\alpha^{i}_{0}\~,\label{jayellone} \\
\ell^{-i}_{II}&\equiv& -\Hf\{q^{i}_{0}, \alpha^{-}_{0}\}
\~\equi{\e{alphaminus}}\~
-\frac{1}{2\alpha^{+}_{0}}\{q^{i}_{0}, L^{\perp}_{0}-a\hbar\}\~,\label{elltwo} \\
E^{-i}&\equiv& \sum_{n\neq 0}\frac{\mathrm{i}}{n}:\alpha^{i}_{-n}\alpha^{-}_{n}: 
\~\equi{\e{alphaminus}}\~ \frac{1}{\alpha^{+}_{0}}E^{i}\~,\label{eeminus}  \\
E^{i}&\equiv&\sum_{n\neq 0}\frac{\mathrm{i}}{n}:\alpha^{i}_{-n}L^{\perp}_{n}: 
\~\equi{\e{normalorder}}\~
\sum_{n>0}\frac{\mathrm{i}}{n}\left(\alpha^{i}_{-n}L^{\perp}_{n}
-L^{\perp}_{-n}\alpha^{i}_{n}\right)
\~\equi{\e{lnalfm}}\~
\sum_{n\neq 0}\frac{\mathrm{i}}{2n}\{\alpha^{i}_{-n},L^{\perp}_{n}\} \cr
&\equi{\e{ccr01}}&\sum_{n\neq 0}\frac{\mathrm{i}}{12n}\sum_{k\in\Z}
\{\alpha^{i}_{-n},\alpha^{j}_{n-k},\alpha^{j^{\prime}}_{k}\}\eta\q_{jj^{\prime}}
\=\sum_{n\neq 0}\frac{\mathrm{i}}{12n}\sum_{\ell\in\Z+\frac{n}{2}}
\{\alpha^{i}_{-n},\alpha^{j}_{\frac{n}{2}-\ell},
\alpha^{j^{\prime}}_{\frac{n}{2}+\ell}\}\eta\q_{jj^{\prime}}\~.\label{defe}
\eea
Note that we have two expressions for the $E^{i}$ operator, either as an
anti-commutator with $L^{\perp}_{n}$, or as a $3$-symmetrizer, which follows
from straightforward manipulations. Hermiticity is manifestly guaranteed by the
anti-commutator ($3$-symmetrizer) form,
\beq
q^{\mu\dagger}_{0}\= q^{\mu}_{0}\~, \qquad 
\alpha^{\mu\dagger}_{n}\=\alpha^{\mu}_{-n}\~, \qquad 
L^{\perp\dagger}_{n}\=L^{\perp}_{-n}\~, \qquad 
J^{\mu\nu\dagger}\=J^{\mu\nu}\~, \qquad
E^{i\dagger}\=E^{i}\~,
\eeq
basically because the anti-commutator ($3$-symmetrizer) of two (three)
Hermitian operators is again Hermitian, respectively.

\section{Commutator $[J^{-i},J^{-j}]$}

\noi
Let us now derive the sought-for commutator
\beq\fbox{$
[J^{-i},J^{-j}] \= \frac{2\hbar^{2}}{(\alpha^{+}_{0})^{2}}\sum_{n\neq 0} 
:\alpha^{i}_{-n} \alpha^{j}_{n}:\kappa^{|n|}\left[n\left(\frac{D-2}{24}-1\right)
-\frac{1}{n}\left(\frac{D-2}{24}-a\right)\right]+{\cal O}(\kappa\!-\!1)\~,$}
\label{jjcom}
\eeq
which, in the limit $\kappa\to 1$, precisely vanishes for $D=26$ and $a=1$

\noi
{\sc Proof of \eq{jjcom}, part 1:} We may assume that the external transverse
indices $i\neq j$ are different (or else the commutator \e{jjcom} vanishes
trivially). Then the operator $\alpha^{i}_{0}$ in the first term of
\eq{jayellone} commutes with everything in the commutator \e{jjcom} (because
it never meets $q^{i}_{0}$), so that one may treat that $\alpha^{i}_{0}$ as a
$c$-number. In particular, the commutator between the two first terms in
\eq{jayellone} vanishes
\beq
[\ell^{-i}_{I},\ell^{-j}_{I}]\~\equi{\e{jayellone}}\~0\~.
\eeq
Also the operators $q^{-}_{0}$ and $\alpha^{+}_{0}$ commute with everything
except each other. This produces the following commutator between the first
term and the two other terms in \eq{jayellone},
\beq
[\ell^{-i}_{I},\ell^{-j}_{II}+E^{-j}]
\=\alpha^{i}_{0} [q^{-}_{0},\frac{1}{\alpha^{+}_{0}}]
\left(-\Hf\{q^{j}_{0}, L^{\perp}_{0}-a\hbar \}\+E^{j}\right) 
\~\equi{\e{ccr01}}\~\frac{i\hbar\alpha^{i}_{0}}{(\alpha^{+}_{0})^{2}}
\left(-\Hf\{q^{j}_{0}, L^{\perp}_{0}-a\hbar \}\+E^{j}\right)\~. \label{ell123}
\eeq
The commutator between the two second terms \e{elltwo} becomes
\bea
[\ell^{-i}_{II},\ell^{-j}_{II}]&\equi{\e{elltwo}}&\frac{1}{(\alpha^{+}_{0})^{2}}
\left[\Hf\{q^{i}_{0}, L^{\perp}_{0}-a\hbar \},
\~\Hf\{q^{j}_{0}, L^{\perp}_{0}-a\hbar\} \right] \cr
&\equi{\e{wcs2}}&\frac{1}{(\alpha^{+}_{0})^{2}}
[L^{\perp}_{0}-a\hbar ,q^{j}_{0}]\Hf\{q^{i}_{0}, L^{\perp}_{0}-a\hbar \}
\-(i\leftrightarrow j) \cr
&\equi{\e{lnalfm}}&-\frac{\mathrm{i}\hbar\alpha^{j}_{0}}{2(\alpha^{+}_{0})^{2}}
\{q^{i}_{0}, L^{\perp}_{0}-a\hbar \}\-(i\leftrightarrow j)\~,
\eea
which cancels against the $[\ell^{-i}_{I},\ell^{-j}_{II}]-(i\leftrightarrow j)$
contribution in \eq{ell123}. In particular, the two center-of-mass parts 
commute
\beq
[\ell^{-i},\ell^{-j}]
\~\equiv\~[\ell^{-i}_{I}+\ell^{-i}_{II},\ell^{-j}_{I}+\ell^{-j}_{II}]\=0\~.
\eeq
Notice that the light-cone Hamiltonian $L^{\perp}_{0}-a\hbar$ commutes with the
operators $E^{j}$ and $E^{ij}$,  
\beq
[L^{\perp}_{0}-a\hbar,E^{j}]\~\equi{\e{lnalfm}}\~0, \qquad\qquad
[L^{\perp}_{0}-a\hbar,E^{ij}]\~\equi{\e{lnalfm}}\~0\~.\label{littlehelp01}
\eeq
Moreover,
\bea
[q^{i}_{0},E^{j}]&\equi{\e{defe}}&\left[q^{i}_{0},
\sum_{n\neq 0}\frac{\mathrm{i}}{2n}\{\alpha^{j}_{-n},L^{\perp}_{n}\}\right]
\~\equi{\e{ccr01}}\~\sum_{n\neq 0}\frac{\mathrm{i}}{2n}\left\{\alpha^{j}_{-n},
[q^{i}_{0},L^{\perp}_{n}]\right\}
\~\equi{\e{lnalfm}}\~
-\sum_{n\neq 0}\frac{\hbar}{2n}\{\alpha^{j}_{-n},\alpha^{i}_{n}\} \cr
&\equi{\e{eemunu}}&\mathrm{i}\hbar E^{ij}\~.\label{littlehelp02}
\eea
Therefore the commutator between the $\ell^{-i}_{II}$ and $E^{-j}$ becomes
\bea
[\ell^{-i}_{II},E^{-j}]
&=&-\frac{1}{2(\alpha^{+}_{0})^{2}}
\left[\{q^{i}_{0}, L^{\perp}_{0}-a\hbar \},E^{j}\right] 
\~\equi{\e{littlehelp01}}\~
-\frac{1}{2(\alpha^{+}_{0})^{2}}\left\{[q^{i}_{0},E^{j}],
L^{\perp}_{0}-a\hbar \right\} \cr
&\equi{\e{littlehelp02}}&-\frac{\mathrm{i}\hbar}{2(\alpha^{+}_{0})^{2}}
\{E^{ij},L^{\perp}_{0}-a\hbar \} 
\~\equi{\e{littlehelp01}}\~
-\frac{\mathrm{i}\hbar}{(\alpha^{+}_{0})^{2}}(L^{\perp}_{0}-a\hbar)E^{ij}
\=-(i\leftrightarrow j) \cr
&\equi{\e{normalhelp}}&
\- \frac{\mathrm{i}\hbar}{(\alpha^{+}_{0})^{2}}:(L^{\perp}_{0}-a\hbar)E^{ij}:
\- \frac{\hbar^{2}}{(\alpha^{+}_{0})^{2}}
\sum_{n\in\Z}\sgn(n)\kappa^{|n|}:\alpha^{i}_{-n}\alpha^{j}_{n}:\~, \label{elle}
\eea
where we in the last equality normal-ordered the expression by using
\bea
\mathrm{i}L^{\perp}_{0}E^{ij} \- :\mathrm{i}L^{\perp}_{0}E^{ij}:
&\equi{\e{eemunu}}&
\sum_{n>0}\frac{1}{n}[L^{\perp}_{0},\alpha^{i}_{-n}]\alpha^{j}_{n}
\- (i\leftrightarrow j)
\~\equi{\e{lnalfm}}\~
\hbar\sum_{n>0}\kappa^{n}\alpha^{i}_{-n}\alpha^{j}_{n}\-(i\leftrightarrow j)\cr
&=&\hbar\sum_{n\in\Z}\sgn(n)\kappa^{|n|}:\alpha^{i}_{-n}\alpha^{j}_{n}:\~.
\label{normalhelp}
\eea
It remains to compute the commutator between two oscillator terms \e{eeminus},
\beq
[E^{-i},E^{-j}]\~\equi{\e{eeminus}}\~
\frac{1}{(\alpha^{+}_{0})^{2}}[E^{i},E^{j}]\~,
\eeq
which we will do in the last Section, \cf \eq{eiejcom}.

\section{Commutator $[E^{i},E^{j}]$ via $3$-symmetrizer}
\label{secee}

\noi
{}Finally, let us derive, \wtho Wick's Theorem, that
\bea
[E^{i},E^{j}]&=&\mathrm{i}\hbar\left(
2:L^{\perp}_{0}E^{ij}:\-\alpha^{i}_{0}E^{j}\+ \alpha^{j}_{0}E^{i}\right) \cr
&&\+ 2\hbar^{2}\sum_{n\neq 0}:\alpha^{i}_{-n} \alpha^{j}_{n}:
\kappa^{|n|}\left[n\left(\frac{D-2}{24}-1\right)+\sgn(n)-\frac{D-2}{24n}\right]
+{\cal O}(\kappa\!-\!1)\~. \label{eiejcom}
\eea
{\sc Proof of \eq{jjcom}, part 2:} If one adds up contributions from
\eq{eiejcom}, \eq{elle}, and the last term in \eq{ell123}, one derives precisely
the $[J^{-i},J^{-j}]$ commutator \e{jjcom}.
\proofbox

\noi
{\sc Proof of \eq{eiejcom}.} Recall that the operator $E^{i}$ is cubic in the
transverse $\alpha$ oscillator modes, \cf \eq{defe}. The fact that the external
transverse indices $i\neq j$ are different implies that there cannot be a
triple commutator term in \eq{wcs3}, nor double contraction terms in \eq{wsn2}.
Thus the commutator
\bea
C^{ij}&\equiv&-[E^{i},E^{j}]
\~\equi{\e{defe}}\~\sum_{m\neq0\neq n}\sum_{k,\ell\in\Z}
\left[\frac{1}{12n}
\{\alpha^{i}_{-n},\alpha^{j^{\prime}}_{n-k},\alpha^{j^{\prime}}_{k}\},\~\frac{1}{12m}
\{\alpha^{j}_{-m},\alpha^{i^{\prime}}_{m-\ell},\alpha^{i^{\prime}}_{\ell}\}\right] \cr
&\equi{\e{wcs3}}&\sum_{m\neq0\neq n}\frac{1}{4mn}\sum_{k,\ell\in\Z}\left(
2[\alpha^{j^{\prime}}_{k},\alpha^{j}_{-m}]
\frac{1}{24}\{\alpha^{i}_{-n},\alpha^{j^{\prime}}_{n-k},
\alpha^{i^{\prime}}_{m-\ell},\alpha^{i^{\prime}}_{\ell}\} \right.\cr
&&\left.\+ 2[\alpha^{j^{\prime}}_{k},\alpha^{i^{\prime}}_{\ell}]
\frac{1}{24}\{\alpha^{i}_{-n},\alpha^{j}_{-m},\alpha^{j^{\prime}}_{n-k},
\alpha^{i^{\prime}}_{m-\ell}\}\right) \- (i\leftrightarrow j) \cr
&\equi{\e{ccr01}}&\sum_{m\neq0\neq n}\frac{\hbar}{2n}\sum_{\ell\in\Z}\kappa^{|m|}
\frac{1}{24}\{\alpha^{i}_{-n},\alpha^{j}_{n-m},
\alpha^{i^{\prime}}_{m-\ell},\alpha^{i^{\prime}}_{\ell}\} \cr
&&+\sum_{m\neq0\neq n}\frac{\hbar}{2mn}\sum_{k\in\Z}k\kappa^{|k|}
\frac{1}{24}\{\alpha^{i}_{-n},\alpha^{j}_{-m},
\alpha^{i^{\prime}}_{n-k},\alpha^{i^{\prime}}_{m+k}\} \- (i\leftrightarrow j) \cr
&\equi{\e{wsn2}}&C^{ij}_{(4)}\+ C^{ij}_{(2)} \label{eecom}
\eea
is a sum of normal ordered terms $C^{ij}_{(4)}$, quartic in the transverse
$\alpha$ oscillator modes; and single contraction terms $C^{ij}_{(2)}$,
quadratic in the transverse $\alpha$ oscillator modes. The single contraction
terms $C^{ij}_{(2)}=C^{ij}_{(2^{\prime})}+C^{ij}_{(2^{\prime\prime})}$ come in two
types. One type $C^{ij}_{(2^{\prime})}$ has a trace over transverse directions, 
\bea
C^{ij}_{(2^{\prime})}&\equiv&
\sum_{m\neq0\neq n}\frac{\hbar}{2n}\sum_{\ell\in\Z}\kappa^{|m|}
c^{i^{\prime}i^{\prime}}_{m-\ell,\ell}:\alpha^{i}_{-n}\alpha^{j}_{n-m}:\cr
&&+\sum_{m\neq0\neq n}\frac{\hbar}{2mn}\sum_{k\in\Z}k\kappa^{|k|}
c^{i^{\prime}i^{\prime}}_{n-k,m+k}:\alpha^{i}_{-n}\alpha^{j}_{-m}:
\-(i\leftrightarrow j)\cr
&\equi{\e{wsn1}}&0 \-\frac{D-2}{4}\sum_{n\neq0}\frac{\hbar^{2}}{n^{2}}
\sum_{k\in\Z}k|n\!-\!k|\kappa^{|k|+|n-k|}:\alpha^{i}_{-n}\alpha^{j}_{n}:
\-(i\leftrightarrow j) \cr
&\equi{\e{am}}&-\frac{D-2}{4}\sum_{n\neq0}\frac{\hbar^{2}}{n^{2}}A_{n}
:\alpha^{i}_{-n}\alpha^{j}_{n}: \-(i\leftrightarrow j) \cr
&\equi{\e{am}}&
\hbar^{2}\frac{D-2}{12}\sum_{n\neq0}\left(\frac{1}{n}-n\right)\kappa^{|n|}
:\alpha^{i}_{-n}\alpha^{j}_{n}:\~,\label{eetransv}
\eea
which becomes proportional to the number $D\!-\!2$ of transverse directions.
The other type $C^{ij}_{(2^{\prime\prime})}$ does not carry a trace over transverse
directions,
\bea
C^{ij}_{(2^{\prime\prime})}&\equiv&
\sum_{m\neq0\neq n}\frac{\hbar}{2n}\sum_{\ell\in\Z}\kappa^{|m|}\left(
2c^{ji^{\prime}}_{n-m,\ell}:\alpha^{i}_{-n}\alpha^{i^{\prime}}_{m-\ell}:
\+ 2c^{ii^{\prime}}_{-n,\ell}:\alpha^{i^{\prime}}_{m-\ell}\alpha^{j}_{n-m}:\right)\cr
&&+\sum_{m\neq0\neq n}\frac{\hbar}{2mn}\sum_{k\in\Z}k\kappa^{|k|}\left(
c^{i^{\prime}j}_{n-k,-m}:\alpha^{i}_{-n}\alpha^{i^{\prime}}_{m+k}:
\+ c^{i^{\prime}j}_{m+k,-m}:\alpha^{i}_{-n}\alpha^{i^{\prime}}_{n-k}: \right.\cr
&&\left.\+ c^{ii^{\prime}}_{-n,m+k}:\alpha^{i^{\prime}}_{n-k}\alpha^{j}_{-m}
\+ c^{ii^{\prime}}_{-n,n-k}:\alpha^{i^{\prime}}_{m+k}\alpha^{j}_{-m}:\right) 
-(i\leftrightarrow j) \cr
&\equi{\e{wsn1}}&\sum_{m\neq0\neq n}\frac{\hbar^{2}}{2n}\kappa^{|m|}\left(
|n\!-\!m|\kappa^{|n-m|}:\alpha^{i}_{-n}\alpha^{j}_{n}:
\+ |n|\kappa^{|n|}:\alpha^{i}_{m-n}\alpha^{j}_{n-m}:\right)\+ 0 \+ 0\cr
&&+\sum_{m\neq0\neq n}\frac{\hbar^{2}}{4mn}(n\!-\!m)\left(
|m|\kappa^{|m|+|n-m|}:\alpha^{i}_{-n}\alpha^{j}_{n}:
\+ |n|\kappa^{|n|+|m-n|}:\alpha^{i}_{m}\alpha^{j}_{-m}: \right)
-(i\leftrightarrow j)\cr
&\stackrel{k=n-m}{=}&\frac{\hbar^{2}}{2}\sum_{k\neq n\neq 0}
\frac{|k|}{n}\kappa^{|k|+|n-k|}:\alpha^{i}_{-n}\alpha^{j}_{n}:
\+ \frac{\hbar^{2}}{2}\sum_{k\neq n\neq 0}
\sgn(n)\kappa^{|n|+|k-n|}:\alpha^{i}_{-k}\alpha^{j}_{k}:\cr
&&\+ \frac{\hbar^{2}}{2}\sum_{m\neq0\neq n}\left(\sgn(m)-\frac{|m|}{n}\right)
\kappa^{|m|+|n-m|}:\alpha^{i}_{-n}\alpha^{j}_{n}:
\- (i\leftrightarrow j) \cr
&=&\frac{\hbar^{2}}{2}\sum_{0\neq k\neq n\neq 0}
\frac{|k|}{n}\kappa^{|k|+|n-k|}:\alpha^{i}_{-n}\alpha^{j}_{n}:
\+ \frac{\hbar^{2}}{2}\sum_{0\neq k\neq n\neq 0}
\sgn(n)\kappa^{|n|+|k-n|}:\alpha^{i}_{-k}\alpha^{j}_{k}:\cr
&&\+\frac{\hbar^{2}}{2}\sum_{0\neq m\neq n\neq 0}
\left(\sgn(m)-\frac{|m|}{n}\right)
\kappa^{|m|+|n-m|}:\alpha^{i}_{-n}\alpha^{j}_{n}:
\- (i\leftrightarrow j) \cr
&\stackrel{n\leftrightarrow k\leftrightarrow m}{=}&
\hbar^{2}\sum_{0\neq k\neq n\neq 0}\sgn(k)\kappa^{|k|+|n-k|}
:\alpha^{i}_{-n}\alpha^{j}_{n}: \- (i\leftrightarrow j) \cr
&\equi{\e{bm}}&\hbar^{2}\sum_{n\neq 0}(B_{n}\!-\!\sgn(n)
\kappa^{|n|}):\alpha^{i}_{-n}\alpha^{j}_{n}:\- (i\leftrightarrow j) \cr
&\equi{\e{bm}}&2\hbar^{2}\sum_{n\neq 0}(n\!-\!\sgn(n))
\kappa^{|n|}:\alpha^{i}_{-n}\alpha^{j}_{n}:\~.\label{eesinglenotr}
\eea
The normal-ordered terms $C^{ij}_{(4)}$ in \eq{eecom} read
\bea
C^{ij}_{(4)}&\equiv&\sum_{m\neq 0\neq n}\frac{\hbar}{2n}\sum_{\ell\in\Z}\kappa^{|m|}
:\alpha^{i}_{-n}\alpha^{j}_{n-m}\alpha^{i^{\prime}}_{m-\ell}\alpha^{i^{\prime}}_{\ell}:
\cr
&&+\sum_{m\neq0\neq n}\frac{\hbar}{2mn}\sum_{k\in\Z}k\kappa^{|k|}
:\alpha^{i}_{-n}\alpha^{j}_{-m}\alpha^{i^{\prime}}_{n-k}\alpha^{i^{\prime}}_{m+k}:
-(i\leftrightarrow j) \cr
&=&\sum_{-k\neq n\neq 0}\frac{\hbar}{2n}\sum_{\ell\in\Z+\frac{n+k}{2}}\kappa^{|n+k|}
:\alpha^{i}_{-n}\alpha^{j}_{-k}
\alpha^{i^{\prime}}_{\frac{n+k}{2}-\ell}\alpha^{i^{\prime}}_{\frac{n+k}{2}+\ell}:\cr
&&+\sum_{m\neq0\neq n}\frac{\hbar}{2mn}
\sum_{\ell\in\Z+\frac{m+n}{2}}(\ell\!+\!\frac{n-m}{2})\kappa^{|\ell+\frac{n-m}{2}|}
:\alpha^{i}_{-n}\alpha^{j}_{-m}
\alpha^{i^{\prime}}_{\frac{m+n}{2}-\ell}\alpha^{i^{\prime}}_{\frac{m+n}{2}+\ell}:
\- (i\leftrightarrow j)\~, \label{eenormal}
\eea
where we in the first term replaced $k=m\!-\!n$ and shifted
$\ell \to\ell\!+\!\frac{n+k}{2}$, while we replaced $k=\ell\!+\!\frac{n-m}{2}$
in the second term. The term in \eq{eenormal} with $\ell$ downstairs is odd
under $\ell \leftrightarrow -\ell$ in the limit $\kappa\to1$, so we can ignore
them from now on. The terms in \eq{eenormal} corresponding to $k=0$ yield
\bea
\sum_{n\neq 0}\frac{\hbar}{2n}\sum_{\ell\in\Z+\frac{n}{2}}\kappa^{|n|}
:\alpha^{i}_{-n}\alpha^{j}_{0}
\alpha^{i^{\prime}}_{\frac{n}{2}-\ell}\alpha^{i^{\prime}}_{\frac{n}{2}+\ell}:
\- (i\leftrightarrow j) 
&\equi{\e{lnperpdef01}}&\sum_{n\neq 0}\frac{\hbar\kappa^{|n|}\alpha^{j}_{0}}{n}
:\alpha^{i}_{-n}L^{\perp}_{n}:\- (i\leftrightarrow j) \cr
&\stackrel{\e{defe}}{\longrightarrow}& 
-\mathrm{i}\hbar\alpha^{j}_{0}E^{i}\- (i\leftrightarrow j)
\~\~\for\~\~\kappa\~\to\~1\~.\label{eenormal01}
\eea
The terms in \eq{eenormal} corresponding to $m\!+\!n\!=\!0$ yield
\beq
-\sum_{n\neq0}\frac{\hbar}{2n}\sum_{\ell\in\Z}\kappa^{|\ell+n|}
:\alpha^{i}_{-n}\alpha^{j}_{n}\alpha^{i^{\prime}}_{-\ell}\alpha^{i^{\prime}}_{\ell}:
\- (i\leftrightarrow j)
\~\longrightarrow\~ -\mathrm{i}\hbar:E^{ij}L^{\perp}_{0}:\- (i\leftrightarrow j)
\~\~\for\~\~\kappa\~\to\~1\~.\label{eenormal02}
\eeq
The remaining terms in \eq{eenormal} vanish
\bea
&&\sum_{0\neq-k\neq n\neq 0}\frac{\hbar}{2n}
\sum_{\ell\in\Z+\frac{n+k}{2}}\kappa^{|n+k|}
:\alpha^{i}_{-n}\alpha^{j}_{-k}
\alpha^{i^{\prime}}_{\frac{n+k}{2}-\ell}\alpha^{i^{\prime}}_{\frac{n+k}{2}+\ell}:\cr
&&+\sum_{0\neq -m\neq n\neq 0}\left(\frac{\hbar}{4m}-\frac{\hbar}{4n}\right)
\sum_{\ell\in\Z+\frac{m+n}{2}}\kappa^{|\ell+\frac{n-m}{2}|}
:\alpha^{i}_{-n}\alpha^{j}_{-m}
\alpha^{i^{\prime}}_{\frac{m+n}{2}-\ell}\alpha^{i^{\prime}}_{\frac{m+n}{2}+\ell}:
\- (i\leftrightarrow j)\cr
&\longrightarrow& 0\~\~\for\~\~\kappa\~\to\~1\~, \label{lasteq}
\eea
which can be seen by renaming $n \leftrightarrow m$ in the term containing
$\frac{\hbar}{4m}$ in \eq{lasteq}. {}Finally, if one adds up contributions from 
eqs.\ \e{eetransv}, \e{eesinglenotr}, \es{eenormal01}{eenormal02}, one 
derives precisely the commutator \e{eiejcom}. 
\proofbox

\vspace{0.8cm}

\noi
{\sc Acknowledgement:}~K.B.\ would like to thank Paul\'ina Karlub\'ikov\'a
for discussions and carefully reading the manuscript. The work of K.B.\ is
supported by the Grant agency of the Czech republic under the grant
P201/12/G028.

\end{document}